\documentstyle[12pt]{article}
\def\simgt{\stackrel{>}{{}_\sim}}
\def\be{\begin{equation}}       
\def\ee{\end{equation}}
\def\bear{\be\begin{array}}      
\def\eear{\end{array}\ee}
\def\bea{\begin{eqnarray}}
\def\eea{\end{eqnarray}}

\def\ie{{\it i.e.}}
\def\etal{{\it et al.}}
\def\half{{\textstyle{1 \over 2}}}

\def\eighth{{\textstyle{1 \over 8}}}
\def\bold#1{\setbox0=\hbox{$#1$}%
     \kern-.025em\copy0\kern-\wd0
     \kern.05em\copy0\kern-\wd0
     \kern-.025em\raise.0433em\box0 }
\begin{document}
\catcode`@=11
\newtoks\@stequation
\def\subequations{\refstepcounter{equation}%
\edef\@savedequation{\the\c@equation}%
  \@stequation=\expandafter{\theequation}
  \edef\@savedtheequation{\the\@stequation}
  \edef\oldtheequation{\theequation}%
  \setcounter{equation}{0}%
  \def\theequation{\oldtheequation\alph{equation}}}
\def\endsubequations{\setcounter{equation}{\@savedequation}%
  \@stequation=\expandafter{\@savedtheequation}%
  \edef\theequation{\the\@stequation}\global\@ignoretrue

\noindent}
\catcode`@=12
\begin{titlepage}

\begin{flushright}
IC/96/189\\
SHEP--96--24\\
hep-ph/9609472\\ 
September 1996
\end{flushright}

\vspace*{5mm}

\begin{center}
{\Huge Chargino Production in Different}\\
\vspace*{5mm}
{\Huge Supergravity Models }\\
\vspace*{5mm}
{\Huge and the Effect of the Tadpoles}\\[15mm]

{\large{\bf Marco Aurelio D\'\i az}\\
\hspace{3cm}\\
{\small High Energy Section, International 
Centre for Theoretical Physics}\\
{\small P. O. Box 586, Trieste 34100, Italy}} 
\end{center}
\vspace{5mm}
\begin{abstract}

If the lightest chargino is discovered at LEP2, the measurement of its 
mass and cross section together with the mass of the lightest neutralino,
enables the determination of the parameters that define the theory and
the entire supersymmetric and Higgs spectrum. Within this context, we:
(i) study the effect of the one--loop tadpoles in the minimization 
condition of the Higgs potential, by comparing the RGE--improved Higgs
potential approximation with the calculation of the minimum including
the effect of the one--loop tadpoles, and (ii) compare the prediction of 
two different supergravity models, namely, the model based on $B=2m_0$
and motivated by the solution of the $\mu$--problem, and the minimal 
supergravity model, based on $A=B+m_0$.

\end{abstract}

\end{titlepage}


\section{Introduction}

Chargino searches at LEP1.5 at $130 {\rm GeV}<\sqrt{s}<140$ GeV center
of mass energy have been negative, and new exclusion regions in the 
$m_{\chi^{\pm}_1}-m_{\chi^0_1}$ plane have been published \cite{130search}.
For example, if $m_{\chi^0_1}=40$ GeV and if $m_{\tilde\nu_e}>200$ GeV, the 
chargino mass satisfy $m_{\chi^{\pm}_1}>65$ GeV at 95 \% C.L. Nevertheless, 
this bound on the chargino mass is relaxed if the lightest neutralino mass
is close to the chargino mass, or if the sneutrino is lighter than 200 GeV.

In Global Supersymmetry with universal gaugino masses, \ie, without assuming 
scalar mass unification
and without imposing radiative breaking of the electroweak symmetry, it 
was shown how the discovery of the lightest chargino at LEP2 and the 
measurement of $m_{\chi^{\pm}_1}$, $\sigma(e^+e^-\longrightarrow\chi^+_1
\chi^-_1)$, and $m_{\chi^0_1}$, will enable the determination of the 
basic parameters $m_{\tilde g}$ (or $M_{1/2}$), $\mu$, and $\tan\beta$
\cite{DiazKingi}. A similar line was follow by ref.~\cite{FengS}, 
without assuming gaugino unification.

The idea of determining the supersymmetric parameters using the measurements
mentioned above, was extended to Supergravity models \cite{DiazKingii}.
The predictive power of these models is greater, thus not only the basic
parameters of the theory can be determined, but also the entire spectrum.
The model analized in ref.~\cite{DiazKingii} satisfies the condition 
$B=2m_0$, where $B$ is the 
bilinear soft mass and $m_0$ is the unified scalar mass. These models are
motivated by the solution of the $\mu$--problem \cite{SUGRAII}. 

In ref.~\cite{DiazKingii}, the radiative breaking of the electroweak 
symmetry was found by minimizing the RGE improved Higgs potential, \ie,
using the tree level condition
\begin{equation}
(m_{1H}^2+\half m_Z^2\cos 2\beta)(1+\cos 2\beta)=
(m_{2H}^2-\half m_Z^2\cos 2\beta)(1-\cos 2\beta)
\label{eq:treeMinimum}
\end{equation}
but with running parameters evaluated at $Q=m_Z^2$. In the above equation, 
$m_{1H}^2$ and $m_{2H}^2$ are the mass parameters of the two Higgs doublet
$H_1$ and $H_2$ respectively, and the effect of the one--loop tadpoles 
have been neglected. 

The purpose of this letter is two folded. First, we include the one--loop
tadpoles and study the effect on the determination of the supersymmetric 
parameters from chargino observables. And second, keeping the one--loop
tadpoles, we study the differences between the supergravity model based
on the relation $B=2m_0$, and the Minimal Supergravity Model, based on
the relation $A=B+m_0$. 

\section{The Effect of Tadpoles} 

At tree level, the tadpole equations which define the minimum of the
Higgs potential are \cite{diazhaberii}
\begin{eqnarray}
t_{01}&=&m_{1H}^2v_{1}-m_{12}^2v_{2}+\eighth(g^2+g'^2)v_{1}
(v_{1}^2-v_{2}^2)\,,\nonumber\\
t_{02}&=&m_{2H}^2v_{2}-m_{12}^2v_{1}+\eighth(g^2+g'^2)v_{2}
(v_{2}^2-v_{1}^2)\,.
\label{eq:tadpoles}
\end{eqnarray}
where $g$ and $g'$ are the gauge coupling constants, $v_1$ and $v_2$ are
the vacuum expectation values of the two Higgs doublets, and $m_{1H}^2$,
$m_{2H}^2$ and $m_{12}^2$ are three arbitrary parameters with units of mass
squared. The minimum of the Higgs potential is defined by 
$t_{01}=t_{02}=0$. Using this latest condition, 
eq.~(\ref{eq:treeMinimum}) can
be derived from eq.~(\ref{eq:tadpoles}) by eliminating $m_{12}^2$ and 
noting that $\tan\beta=v_2/v_1$.

At one--loop level, and working with Dimensional Reduction (DRED) in the
$\overline{MS}$ scheme, the tree level quantities depend implicitly  now 
on the arbitrary scale $Q$. On the other hand, one--loop contributions to
the tadpoles depend explicitly on that scale, such that the renormalized
tadpoles are scale independent at the one--loop level:
\begin{eqnarray}                  
t_1&=&\left[m_{1H}^2v_{1}-m_{12}^2v_{2}+                                
\eighth(g^2+g'^2)v_{1}(v_{1}^2-v_{2}^2)\right](Q)+                   
\widetilde{T}_1^{\overline{MS}}(Q)                        
\,,\nonumber\\                                           
t_2&=&\left[m_{2H}^2v_{2}-m_{12}^2v_{1}+                                
\eighth(g^2+g'^2)v_{2}(v_{2}^2-v_{1}^2)\right](Q)+                      
\widetilde{T}_2^{\overline{MS}}(Q)\,.                             
\label{eq:RunTad}                                                 
\end{eqnarray}                                            
In the last equation, $\widetilde T_i^{\overline{MS}}(Q)$ are the 
renormalized one--loop tadpoles, \ie, the counterterms have subtracted 
the infinite pieces.

Imposing that the renormalized tadpoles are equal to zero and eliminating
the running parameter $m_{12}^2(Q)$, we find the corrected minimization
condition
\begin{equation}
\left[m_{1H}^2+{\textstyle{1\over{v_1}}}\widetilde T_1^{\overline{MS}}
+\half m_Z^2c_{2\beta}\right]c_{\beta}^2=
\left[m_{2H}^2+{\textstyle{1\over{v_2}}}\widetilde T_2^{\overline{MS}}
-\half m_Z^2c_{2\beta}\right]s_{\beta}^2
\label{eq:OneLoopMin}
\end{equation}
where the dependence on the arbitrary scale $Q$ has been omitted from all
the parameters and tadpoles. Here we calculate the one--loop tadpoles
including loops with top and bottom quarks and squarks.

The difference between including or not including the one--loop tadpoles 
into the minimization condition can be appreciated in Fig.~1 and 2. In 
Fig.~1 we plot the total light chargino pair production cross section as
a function of the lightest neutralino mass. Three scenarios are considered
according to the value of the chargino and gluino masses: $(m_{\chi^{\pm}_1}, 
m_{\tilde g})=(70, 320)$, $(80, 350)$, and $(90, 390)$ GeV. The 
calculation of the radiative breaking of the electroweak symmetry without
the one--loop tadpoles [Eq.~(\ref{eq:treeMinimum})] is represented by 
dashed lines, and the same calculation with one--loop tadpoles 
[Eq.~(\ref{eq:OneLoopMin})] is represented with solid lines. We can 
appreciate that the RGE--improved Higgs potential is not a bad 
approximation in this light chargino scenario, introducing an error typically
of 0.1 pb (3\%) in the cross section, and 0.5 GeV (1\%) in the neutralino 
mass in the most disfavored regions.

In Fig.~2 we plot the total light chargino production cross section as a 
function of (a) the universal scalar mass $m_0$, (b) the trilinear coupling
$A$, (c) the ratio of the two vacuum expectation values $\tan\beta$, and
(d) the supersymmetric Higgs mass parameter $\mu$, for the same cases as
in Fig.~1. With Fig.~2 we see how the fundamental parameters of the 
theory can be determined from the discovery of the chargino and the 
measurement of its mass and cross section, together with the mass of its 
decay product, the lightest neutralino mass. It is clear from the figure 
that the RGE--improved Higgs potential in the light chargino scenario is 
better in the chargino/neutralino sector where the errors in the 
determination of the parameters $\tan\beta$ and $\mu$ are small 
(Figs.~1c and 1d). The
exception occurs in the large $\tan\beta$ region: when the one--loop 
tadpoles are included, large values of $\tan\beta$ are allowed resulting 
in longer (solid) curves. Errors in the determination of $m_0$ and $A$
are larger (Figs.~1a and 1b) what implies the introduction of larger 
errors in the determination of sfermion masses (typically 5--10 \% in the 
disfavored regions).

\section{Two Supergravity Models}

Supergravity models considered here involve the following universal soft 
parameters: $m_0$, $M_{1/2}$, $A$, and $B$. 
Our input parameters are (i) the gluino mass 
$m_{\tilde g}$, which fixes the value of $M_{1/2}$, (ii) $\mu$, and (iii) 
the lightest chargino mass $m_{\chi^{\pm}_1}$, which together with $\mu$ 
fixes the value of $\tan\beta$. The value of $m_0$ is chosen such that 
the minimization condition in eq.~(\ref{eq:OneLoopMin}) is satisfied.
Similarly, up to now, the value of $A$ has been chosen such that the 
relation $B=2m_0$ is satisfied.

An immediate question arises: how do the predictions change if we consider a 
different supergravity model? For this purpose we compare the previous
model ($B=2m_0$) with Minimal Supergravity, where the boundary condition
at the unification scale is $A=B+m_0$.

In Fig.~3 we plot the total light chargino pair production cross section
as a function of the lightest neutralino mass. The values of the light
chargino and gluino masses are the same as in Fig.~1. The model $B=2m_0$
is represented with solid lines and minimal supergravity, with $A=B+m_0$,
is represented with dashed lines. In the case of $B=2m_0$ the sign of 
$\mu$ is unambiguous (positive, in our convention) because the product 
$B\mu$ is related to the CP--odd Higgs mass $m_A^2$. On the contrary, in 
minimal supergravity, $\mu$ can take either sign, nevertheless, in 
Fig.~3 only $\mu>0$ is present because it is not possible to find a 
solution with $\mu<0$ for the displayed choices of $m_{\chi^{\pm}_1}$ and 
$m_{\tilde g}$. Lighter values of $m_{\tilde g}$ are needed in order to 
find a solution with $\mu<0$. We do not show them.
In the three cases presented here, the two models coincide at the upper left
corner of the curves; this is because in those cases $A\approx3m_0$, and
both type of boundary condition are satisfied simultaneously. This occurs
at small values of $\tan\beta$ ($\sim 2$) and large values of $\mu$ 
($\sim 300$ GeV). We can notice also that in these models the lightest
neutralino mass satisfy $m_{\chi^0_1}\simgt\half m_{\chi^{\pm}_1}$,
which is important for chargino searches, considering that the bounds
are not applicable if $m_{\chi^{\pm}_1}-m_{\chi^0_1}<10$ GeV.

The differences between the two models can be also appreciated in the 
prediction of the masses of the supersymmetric partners, as it is 
displayed in the next figure. In Fig.~4 we plot the total light
chargino pair production cross section as a function of (a) the second 
lightest neutralino mass $m_{\chi^0_2}$, (b) the sneutrino mass 
$m_{\tilde\nu}$, which for all practical purposes is degenerate for 
the three flavors, (c) the lightest charged slepton mass 
$m_{\tilde l^{\pm}_1}$, which is always the stau, and (d) the lightest
up--type squark mass $m_{\tilde q_{u1}}$, which is mostly stop. 
The prediction of the $\chi^0_2$ mass is quite similar in both models,
as it can be appreciated from Fig.~4a, and this mass satisy
$m_{\chi^0_2}\approx m_{\chi^{\pm}_1}$. On the contrary, differences are more
pronounced in the sfermion sector, and the general trend is that 
sneutrinos, staus and stops are lighter in the $B=2m_0$ model 
compared with Minimal Supergravity (Figs.~4b--4d).
It is worth to notice that in the light chargino scenario, most of the 
time the sneutrino mass is smaller than 200 GeV (Fig.~4b). This is 
important for chargino searches, because it is common to assume a sneutrino
heavier that 200 GeV in quoting chargino mass bounds. 

\section{Conclusions}

The discovery of a light chargino at LEP2 and the measurement of its mass 
and cross section, as well as the mass of the lightest neutralino, 
enable the determination of the basic parameters of the theory and,
in supergravity models, the prediction of the entire supersymmetric and 
Higgs spectrum. Within this light chargino context, we have shown that the 
RGE improved Higgs potential 
approximation, \ie, the omision of the one--loop tadpoles contributions,
introduce an error of a few percent in the determination of the 
parameters and masses of the chargino/neutralino sector. With the 
exception of large $\tan\beta$ region, where errors are larger. 
The error is also larger in the sfermion sector (typically of 10\%). 

At the same time, we have shown how the predictions change in different
supergravity models. We compared a model based on the boundary 
condition $B=2m_0$ and motivated 
by the solution of the $\mu$--problem, with Minimal Supergravity based on
the boundary condition $A=B+m_0$. When the same sign of $\mu$ is considered,
predictions are quite different except in the large $\tan\beta$ region 
where the two models tend to coincide ($A\approx3m_0$). Nevertheless, 
minimal supergravity
accepts the other sign of $\mu$, and that class of solutions has no 
counterpart in the $B=2m_0$ model.

\section*{Acknowledgements}

The author is indebted to Prof. Steve King for his input in the subject.
Part of this work was done in the University of Southampton, 
Southampton, England.

\section*{Figure Captions}

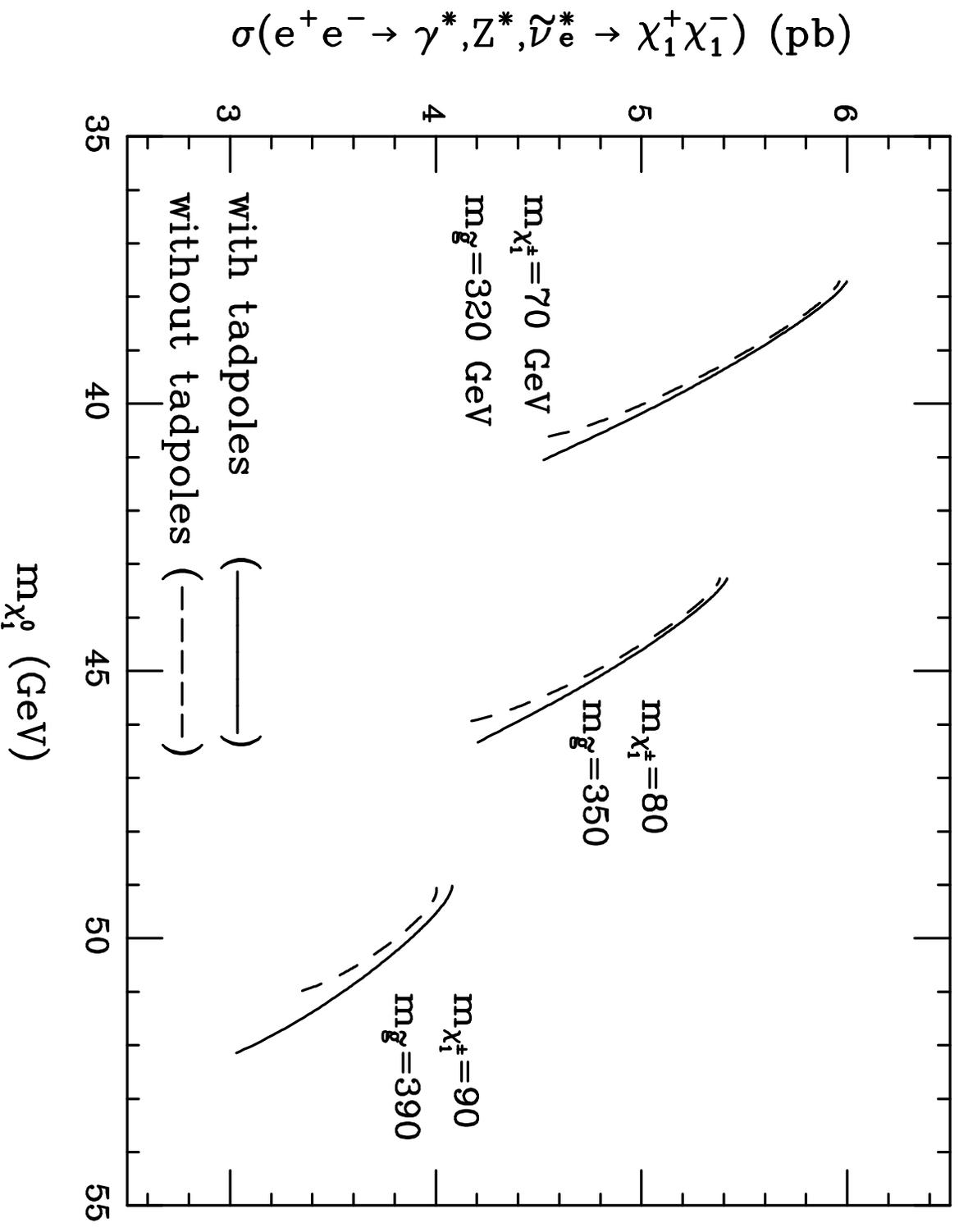
\noindent {\bf Fig. 1:} 
Total light chargino pair production cross section as a function of
the lightest neutralino mass for three different values of the chargino
and gluino masses: $(m_{\chi^{\pm}_1}, m_{\tilde g})=(70, 320)$, 
$(80, 350)$, and $(90, 390)$ GeV. We take the boundary condition at the
unification scale to be $B=2m_0$. The dashed lines are found neglecting
the effect of the tadpoles, while the solid lines include this effect.
\vskip 0.3cm

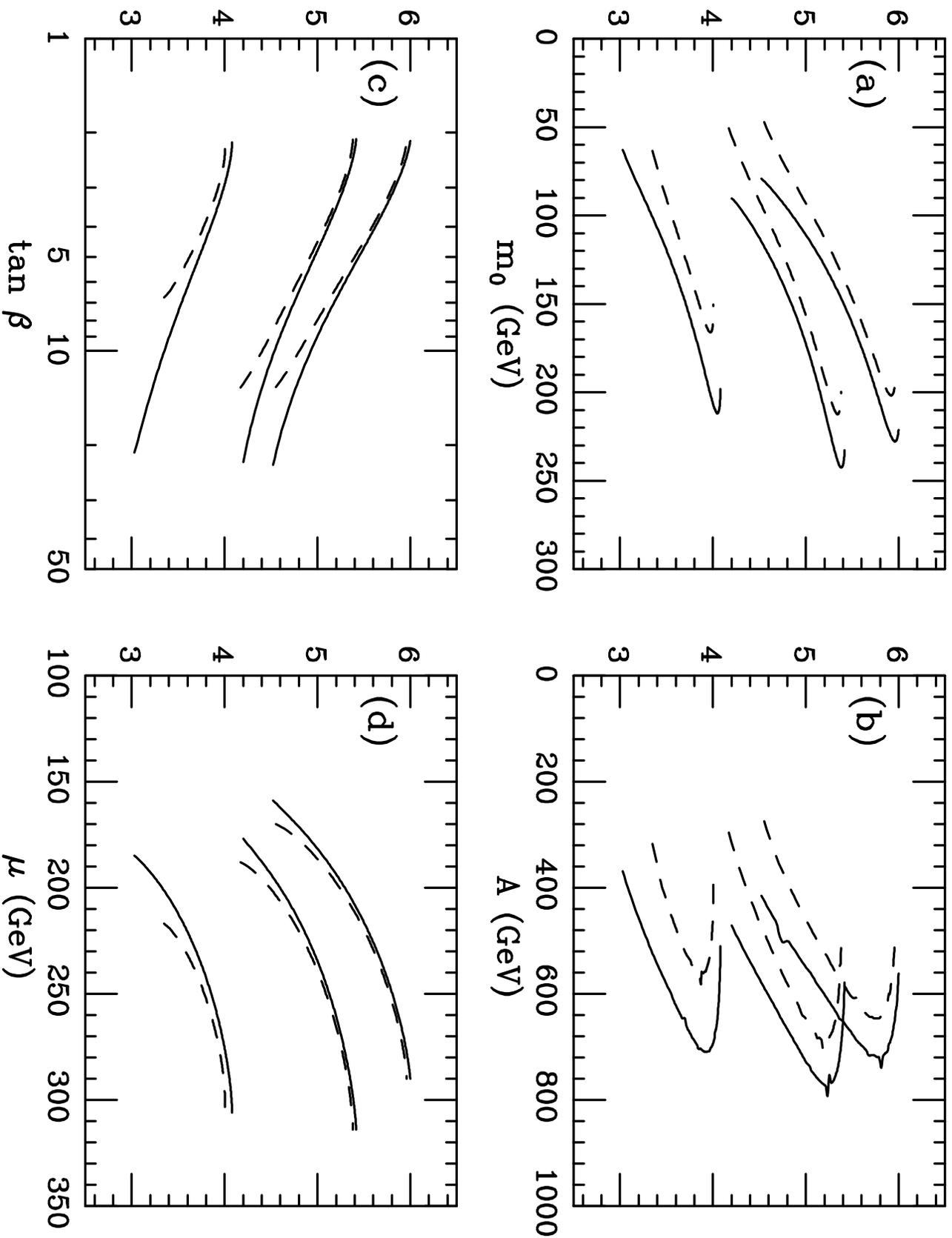
\noindent {\bf Fig. 2:}
For the same cases as in Fig. 1, the total light chargino pair production 
cross section is plotted as a function of (a) $m_0$, (b) $A$, (c) 
$\tan\beta$, and (d) $\mu$. Solid lines include one--loop tadpoles and
dashed lines do not. 
\vskip 0.3cm

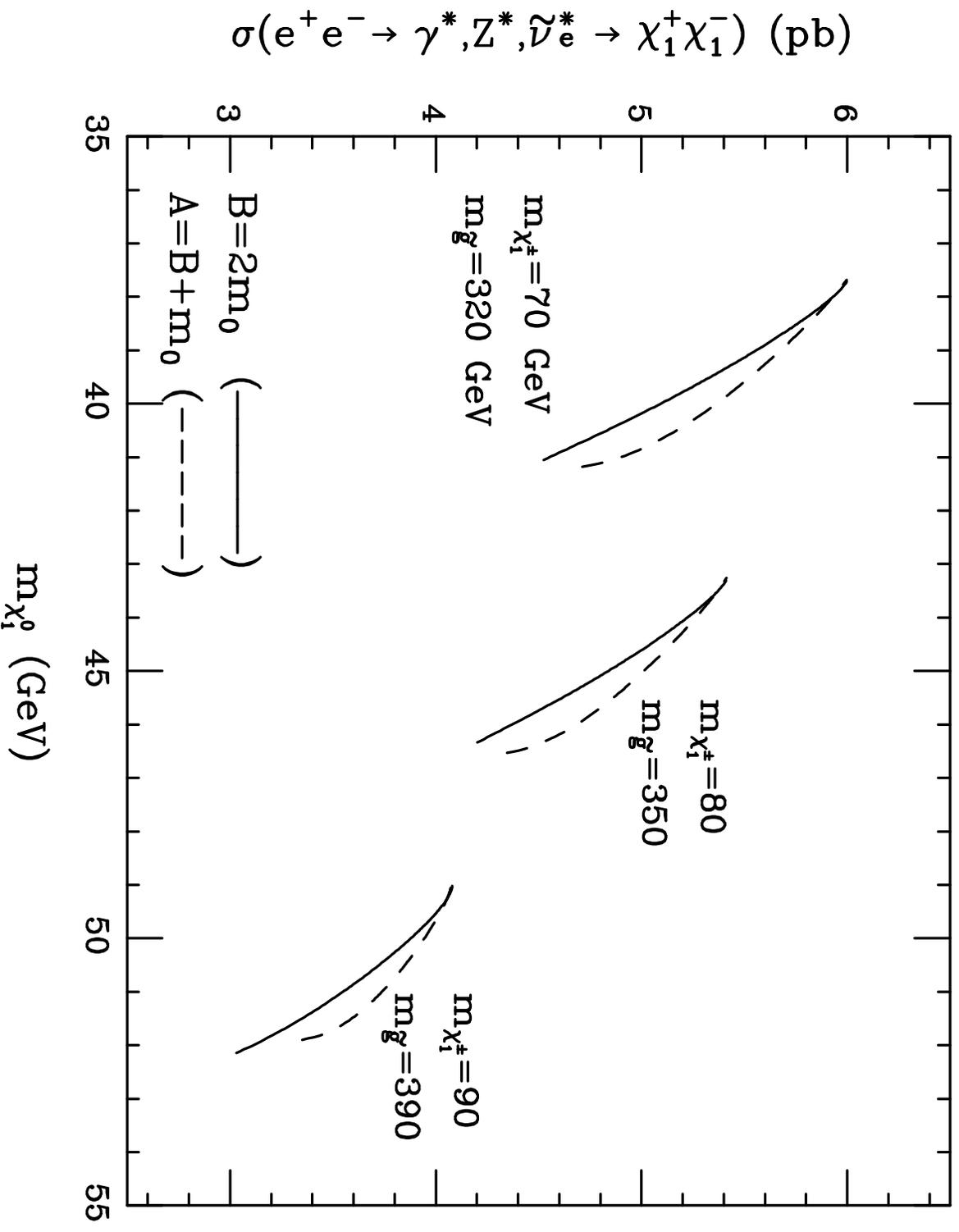
\noindent {\bf Fig. 3:}
Total light chargino pair production cross section as a function of
the lightest neutralino mass for three different values of the chargino
and gluino masses: $(m_{\chi^{\pm}_1}, m_{\tilde g})=(70, 320)$, 
$(80, 350)$, and $(90, 390)$ GeV (same as in Fig. 1). In solid lines, the
$B=2m_0$ boundary condition is used, while $A=B+m_0$ is used in the 
dashed lines. In all curves the effect of one--loop tadpoles is included.
\vskip 0.3cm

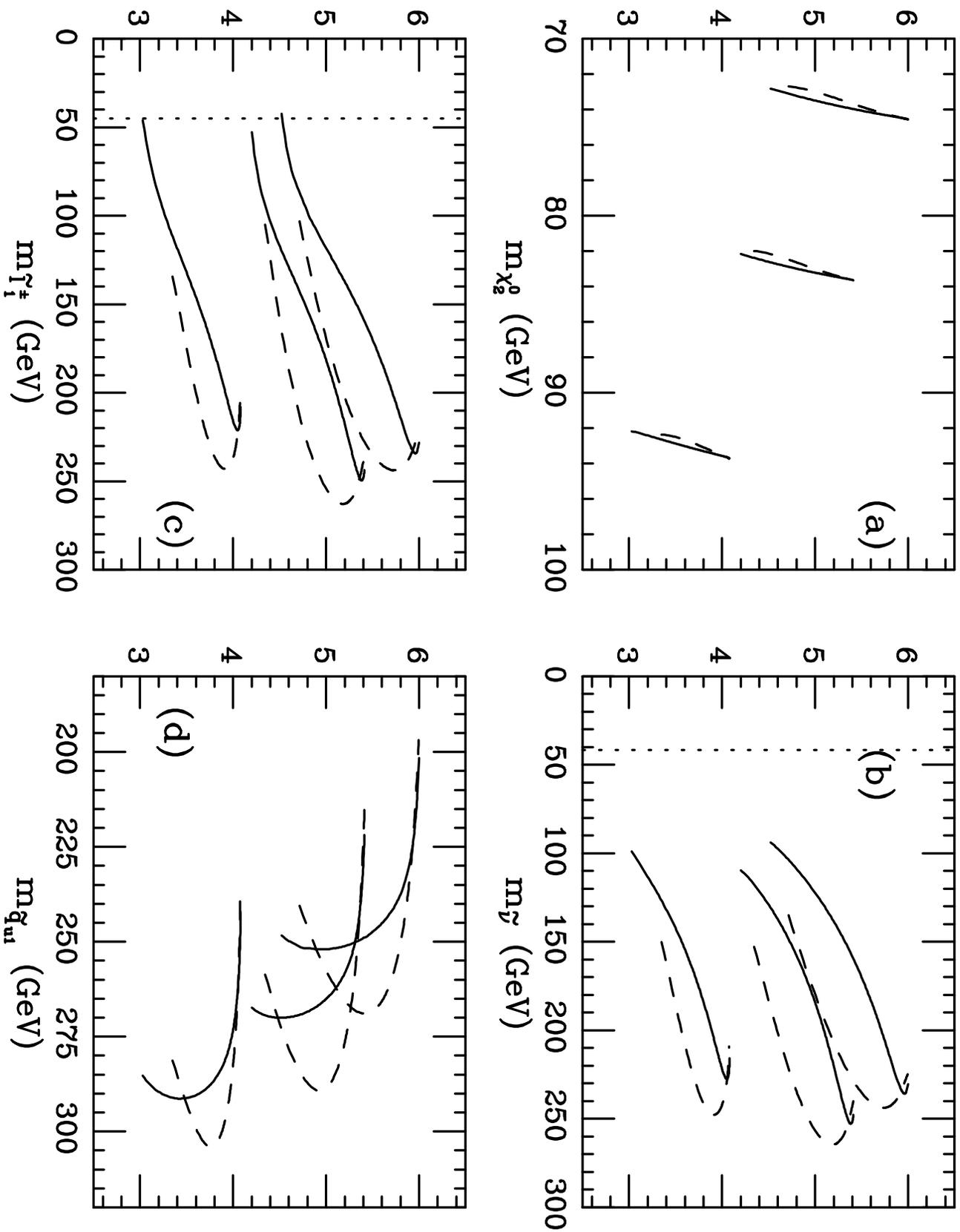
\noindent {\bf Fig. 4:}
For the same cases as in Fig. 3, the total light chargino pair production 
cross section is plotted as a function of (a) $m_{\chi^0_2}$, (b) 
$m_{\tilde\nu}$, (c) $m_{\tilde l_1^{\pm}}$, and (d) $m_{\tilde q_{u1}}$. 
Solid lines correspond to $B=2m_0$ and dashed lines correspond to 
$A=B+m_0$. All curves include the effect of one--loop tadpoles.
\vskip 0.3cm

\end{document}